\documentclass[twocolumn,showpacs,preprintnumbers,amsmath,amssymb]{revtex4}
\usepackage[dvips]{graphicx} 
\begin{document}

\preprint{IFF-RCA-05-10}
\title{Quantum state of the multiverse}
\author{Salvador Robles-P\'{e}rez and Pedro F. Gonz\'alez-D\'{\i}az}
\affiliation{Colina de los Chopos, Centro de F\'{\i}sica ``Miguel Catal\'{a}n'',
Instituto de F\'{\i}sica Fundamental,
Consejo Superior de Investigaciones Cient\'{\i}ficas, Serrano 121, 28006
Madrid (SPAIN) and \\ Estaci\'{o}n Ecol\'{o}gica de Biocosmolog\'{\i}a, Pedro de Alvarado, 14, 06411-Medell\'{\i}n, (SPAIN).}
\date{\today}

\begin{abstract}
A third quantization formalism is applied to a simplified multiverse
scenario. A well defined quantum state of the multiverse is obtained
which agrees with standard boundary condition proposals. These states
are found to be squeezed, and related to accelerating universes: they
share similar properties to those obtained previously by Grishchuk and
Siderov. We also comment on related works that have criticized the
third quantization approach.
\end{abstract}


\maketitle

\section{Introduction}

It is usually claimed that the quantum state of the universe is
described by a wavefunction \cite{Hartle83} which should account for all the
physical information which can be extracted from the universe.
Two main approaches have been followed in order to obtain such a
wavefunction of the universe. First, it can be obtained as the
solution of the Wheeler-De Witt equation \cite{DeWitt67}. This can actually be
seen as a Schr\"{o}dinger equation in which no time
derivative appears in order to satisfy the time invariance required by the
general relativity theory \cite{Hawking81}. Secondly, a path integral approach can
also be taken \cite{Hartle83}. Then, the wavefunction of the universe is given by a sum over all geometries and field
configurations which can be matched with a
given value of the field configuration, defined in a
spatial section of the whole space-time manifold. We make a Wick rotation to Euclidean time in
order to obtain well-defined integrals, and then rotate back to Lorentzian time to obtain the final
results.

In both cases, some boundary conditions have to be imposed \cite{Hartle83, Vilenkin86}. The Hartle-Hawking
no-boundary proposal \cite{Hartle83} represents a universe which is
created from "nothing", meaning by that the absence
of space, time and matter. Vilenkin \cite{Vilenkin82} considered a
universe which is also created from nothing, but through a quantum
tunneling transition instead. Vilenkin argued that the idea of a
universe being created from nothing is not crazier than the creation of particle in other quantum theories.
Thus, it seems appropriate to
consider that the universe may be created as a quantum
fluctuation of the gravitational vacuum. Once a universe nucleates
in the space-time foam \cite{Wheeler57, Hawking78}, it may bubble and eventually jump into an inflationary
period \cite{Vilenkin82}, which is supposed to be the origin of our
current universe. Therefore, we must consider a multiverse in which changes in the topology of space-time
are allowed.

A variety of multiverse hypotheses have been recently considered from different cosmological
viewpoints \cite{Barrau07}. The specific meaning which is ascribed to a single universe
depends on the formalism of the relevant theory. Some well-known examples include the
Everett's many-world interpretation of the quantum theory \cite{Everett57},
the chaotic inflationary multiverse \cite{Linde86} or the landscape in string
theory \cite{Davies04}, among others (for an exhaustive review, see Ref. \cite{Carr07} and references therein). In this paper we only consider the case of a multiverse which is described by a set of quantum oscillators, each one representing a causally disconnected region of the whole space-time.

Topology changes were first claimed to appear in the quantum physics of black
hole evaporation \cite{Hawking90}, and so first arose as a result of trying to
take quantum physics seriously as a description of the whole universe.
On pure cosmological grounds, the transition from a matter dominated
universe into a space-time filled with phantom energy provides us
with another example of a bifurcating topology originated from the
big rip singularity which splits space-time into two disconnected
regions.

Creation of universes and therefore topology changes are naturally contemplated in the third quantization
formalism \cite{Strominger90}. This consists of a further quantization of the
wavefunction of the universe similarly to how quantum field theory
is constructed from the Schr\"{o}dinger wavefunction of matter
fields. The computations are difficult to perform in
the general case. In this paper, however, a simplified model will be
presented in which such computations can in fact be
carried out.

Moreover, the general scheme of the third quantization formalism
can be applied to both, a multiverse made up of parent universes
and a space-time foam filled with virtual baby universes \cite{Strominger90}. Parent
universes are defined to be large space-time regions with a
Hubble length of the same order as the Hubble length of our universe. Baby universes are
considered to be virtual fluctuations of the metric in the vacuum
of gravity, and their contribution to calculations of the gravitational field is extremely important at
the Planck length. The gravitational vacuum would be also populated with virtual Lorentzian and
Euclidean wormholes.  The former can be seen as solutions of the Einstein's equations with at least two
asymptotically flat regions, connecting two separate parts either of
the same universe or of two different universes. The latter can be considered as
Euclidean sectors of a Friedman space-time \cite{Hawking90, Hawking88}, whose quantum states
can be seen as the
exponentially decaying versions of the oscillatory universes from
which they were Wick rotated.

Furthermore, the vacuum state of gravity might also be observable
as far as it could induce a loss of quantum coherence in the
matter fields. The debate was centered (see e.g. Refs.
\cite{Coleman88} and \cite{GonzalezDiaz92}) around the assumption
of taking doubly or even multiply connected wormholes instead of
simply connected ones in the quantum vacuum of gravity.
Equivalently, it can be placed in terms of whether virtual baby
universes are created as single universes or in pairs. In the
theory of the quantum multiverse, in which it has been shown that
general topological changes may occur, the loss of quantum
coherence in the matter field sector seems to be unavoidable.

Our aim in this paper is then to apply the third quantization formalism
given in Ref. \cite{Strominger90} to a simplified model of the universe, which
nevertheless retains the fundamental features of the quantum
theory when it is applied to the universe as a whole. It will be
shown that in such a model a well-defined quantum state for the
multiverse can be obtained, which satisfies the usual boundary
conditions. Squeezed states \cite{Walls83}, which are usually
interpreted as quantum states without any classical analog \cite{Walls83, Reid86}, are
found in the context of the multiverse, and in particular they
appear in the context of accelerated universes. This seems to give support to the idea
that the acceleration is due to distinctively quantum mechanical
effects \cite{RoblesPerez08, GonzalezDiaz09}. Furthermore, the
third quantization formalism used in this paper shows some
advantages respect to other approaches which are usually taken in
quantum cosmology. For instance, it can be demonstrated that the
quantum state of the multiverse may be given in terms of the
states of a quantum harmonic oscillator. Dealing with frequencies
instead of potentials simplifies the computations. Moreover, neither
a perturbative nor a semiclassical approximation needs to be taken
in the model presented in this paper, and therefore the state of
the multiverse which is obtained can describe the global state of
the multiverse as a whole.

Before going any further, however, some caveats and comments should
be made about the ideas and results considered in this paper.
First of all, dealing with squeezed states in the context of the quantum
state of the multiverse would, at first sight, seem somewhat extraneous.
Actually, squeezed states can be readily interpreted both in quantum optics
and in a space-time foam made up of baby universes, which are both deployed
in a common space-time. Baby universes can in fact be taken as tiny particles
representing small perturbations of the space-time field, whose quantum state
affects the vacuum state of the matter field sector, and so this theory may
turn out to be testable.

It is more difficult to imagine how we could test a theory of a
multiverse made up of parent universes. It looks quite different
even though wormhole communications channels may exist through
which microscopic particles could travel from one universe to
other \cite{GonzalezDiaz07b}. Furthermore, some correlations
among the quantum state of different universes could also be
considered. For instance, if the creation of universes would be
produced in entangled pairs, whose correlations might induce some
observable effects on the state of each individual universe in the
pair \cite{Rozas08}.

Squeezed states have already been studied in other cosmological contexts
such as the inflationary universe and gravitational waves \cite{Grishchuk90, Massimo94}.
It was shown \cite{Grishchuk90} that the gravitational vacuum is populated by gravitons
which evolve into a highly squeezed state. In this paper it is also demonstrated in the
context of a third quantization formalism that a space-time foam made up of baby universes
is in a squeezed state. If such a result could be extrapolated to parent universes, these
universes might not be independent but quantum mechanically correlated, too.

The paper can then be outlined as follows: A discussion with the
main features of the third quantization formalism has been also
added as Sec. II. In Sec. III, it is obtained the wavefunction for
a quantum multiverse made up of Friedmann space-times filled with
an homogeneous and isotropic fluid, and the appropriate Fock space
is defined. We then analyze the two interesting limits of the
quantum state of both a large parent universe and the quantum
gravitational vacuum, which turn out to be described by a squeezed
state with no classical analog. In Sec. IV, the quantum state of
the multiverse is represented by a density matrix rather than by a
wavefunction, accounting thus for mixed states as well as pure
ones. Three representations are employed: first, the second
quantized wavefunction is used as the configuration variable, the
usual boundary conditions of Refs. \cite{Hartle83, Vilenkin86} are
applied and a probability interpretation can then be used.
Secondly, a general squeezed number representation is taken in
order to study the quantum state of the multiverse. Parent
universes will be represented by number states. We show that the
quantum state for the vacuum of baby universes is given by a
squeezed state, and that high order correlations appear among
them. For the sake of completeness, we have also used the usual
$P(\alpha)$ representation in terms of coherent states. In Sec. V,
we compare the results obtained in this paper with previous works
which have considered or criticized the third quantization
formalism. The conclusions are collected in Sec. VI, together with
some comments to extend the model to a two-dimensional wave equation which would explicitly account for other matter fields.

\section{The third quantization formalism}

\begin{figure}[h]

\begin{center}

\includegraphics[width=8cm]{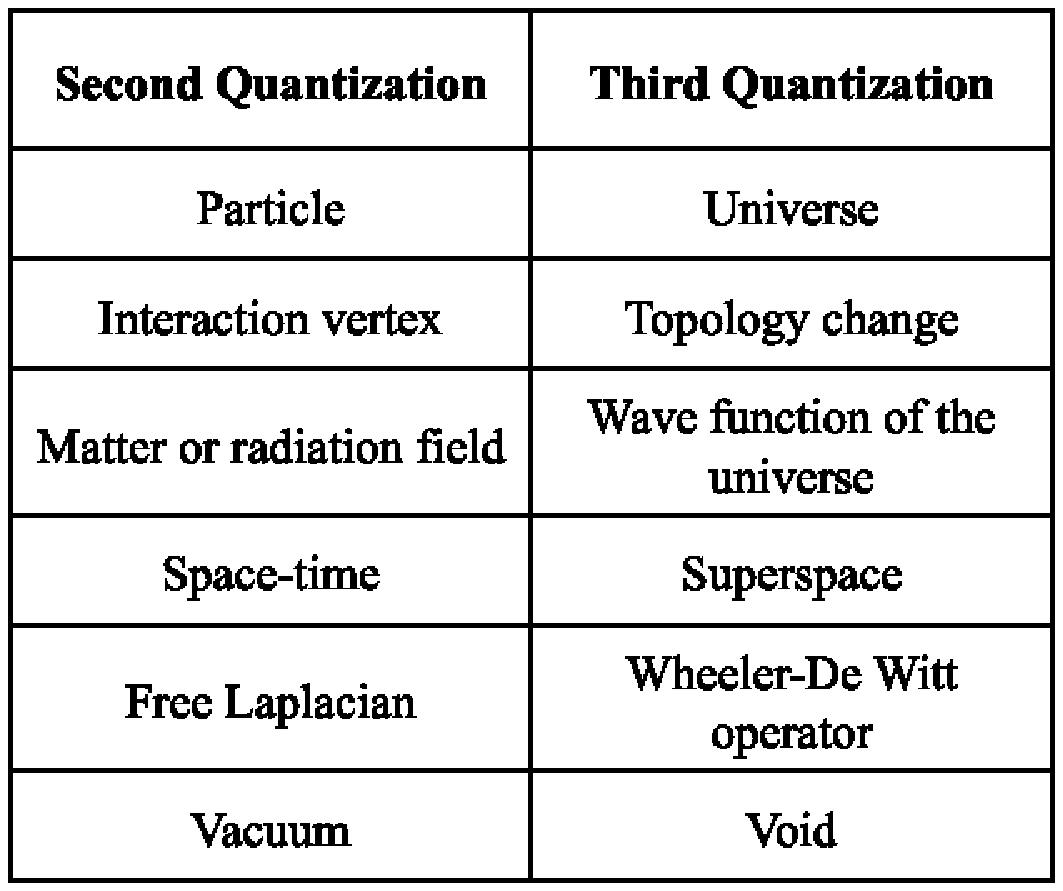}

\end{center}

\caption{Analogy between the second and the third quantization schemes \cite{Strominger90}.}

\label{comparative}

\end{figure}

The basic idea of the third quantization formalism is  \cite{Strominger90}:
to treat the many-universe system as a quantum field theory on superspace.
However, the name 'third quantization' may be misleading. Actually, the procedure
is formally similar to that of the 'second quantization' used in the quantum field theory
of matter fields (see, Fig. \ref{comparative}). The name third quantization comes from the fact that the field which is
quantized is the wave function of the universe, and it depends on \emph{already} second
quantized matter fields, i.e. it depends on the particles existing within the universe.

Broadly speaking, the third quantization procedure starts from the Wheeler-De Witt equation,
it takes one of the fields in that equation as a time variable and then we can proceed as
it is usually done in quantum field theory. However, it not clear at all that such a
'time variable field' can be found in the general 3+1 dimensional theory. Such a
field can be found in one-dimensional cosmological models as well as in 3+1 dimensional models
with high symmetry. Therefore, let us restrict our attention in this section to those cases.

\subsection{One dimensional universes}

Let us start then with one-dimensional model (1 temporal + 0 spatial dimensions).
Throughout this section, we are closely following Ref. \cite{Strominger90}. In that model,
a universe is represented by a point in space. However, if topology changes are allowed,
a many-universe system is rather described by a set of points. On the other hand,
time dependent matter fields within the universes are contemplated. Therefore, let us
start with the second quantized action for such a system. It reads,
\begin{equation}\label{A1}
S = \int_{\tau_0}^{\tau_f} d\tau \left( \frac{1}{\mathcal{N}} G_{\mu\nu} \dot{X}^\mu \dot{X}^\nu - \mathcal{N} m^2 \right) ,
\end{equation}
where $\dot{X}^\mu$, $\mu = 1,..., D$ represent $D$ matter fields, $G_{\mu\nu}$ is the
supermetric in that superspace and $\mathcal{N}$ is the lapse function. For simplicity, a mass term
has been chosen to be the potential of the fields in the action (\ref{A1}). Then, we can
proceed as usual by defining the conjugate momenta of the fields,
$P_\mu = \frac{1}{\mathcal{N}} G_{\mu\nu} \dot{X}^\nu$, and the Hamiltonian density of the system, i.e.
\begin{equation}\label{A2}
H = G^{\mu\nu} P_\mu P_\nu + m^2 .
\end{equation}
In the canonical quantization procedure, $\hat{P}_\mu \rightarrow -i\hbar \frac{\partial}{\partial X^\mu}$,
and the classical Hamiltonian constraint, $H = 0$, turns out to be, $\hat{H}\phi=0$.
This is the Wheeler-De Witt equation for the model being considered and then,
$\phi\equiv \phi(X^\mu)$, represents the (second quantized) wave function of a single universe.

To third quantize this theory, let us find an action whose variational principle leads
to the Wheeler-De Witt equation. This can be done with the following third quantized action,
\begin{eqnarray}\nonumber
^3S &=& \frac{1}{2} \int d^D X \sqrt{-G} \,\, \phi H \phi \\ \label{A3} &=& \frac{1}{2} \int d^D X \sqrt{-G} \left( G^{\mu\nu} \nabla_\mu \phi \nabla_\nu \phi + m^2 \phi^2 \right) ,
\end{eqnarray}
where $G\equiv \rm{det} \it{G}_{\mu\nu}$. Variation of the action (\ref{A3}) with respect
to $\phi$ leads directly to the Wheeler-De Witt equation and, therefore, all the information
of the second quantized theory is preserved in the third quantization formalism \cite{Strominger90}.

\begin{figure}[h]

\begin{center}

\includegraphics[width=8cm]{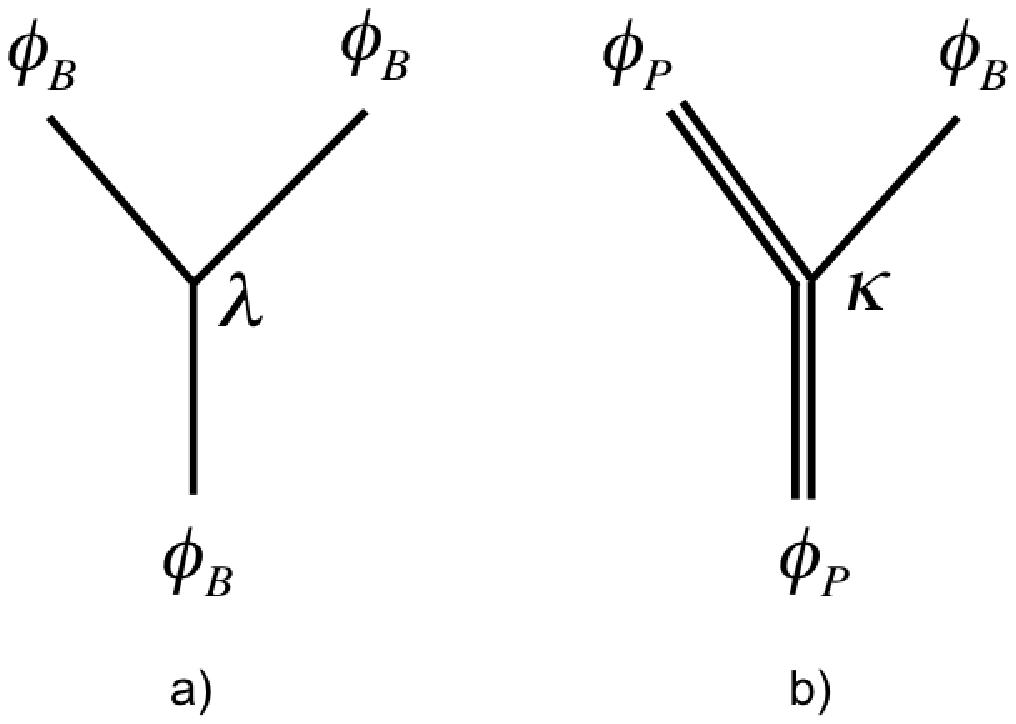}

\end{center}

\caption{Single and doubled lines represent, respectively, the propagators for baby and parent universes: a) the bifurcation of a baby universe; b) the nucleation of a baby universe in the parent space-time.}

\label{bifurcation}

\end{figure}

The usual formalism of quantum field theory can then be directly applied to the third
quantization many-universe system. For instance, the three point Euclidean Green function
can be defined as \cite{Strominger90},
\begin{widetext}
\begin{eqnarray}\label{A4}
G_E(X^\mu_1, X^\mu_2, X^\mu_3) = - \lambda \int d^D X_0^\mu \sqrt{G} \, \, G_E(X_1^\mu, X_0^\mu) G_E(X_2^\mu, X_0^\mu) G_E(X_3^\mu, X_0^\mu).
\end{eqnarray}
\end{widetext}
$\lambda$ is the coupling constant and $G_E(X_i^\mu, X_0^\mu)$ is the two point Euclidean Green function,
i.e.
\begin{equation}
G_E(X_i^\mu, X_f^\mu) = \int \mathcal{D}\phi \,\, \phi(X_i^\mu) \phi(X_f^\mu) e^{- S_E[\phi]}.
\end{equation}
$S_E[\phi]$ is the Euclidean version of the third quantized action (\ref{A3}) for this
interaction scheme, i.e.
\begin{equation}
S_E = \frac{1}{2} \int d^DX^\mu \sqrt{G} \left( G^{\mu\nu} \nabla_\mu \phi \nabla_\nu \phi + m^2 \phi^2 + \frac{\lambda}{3} \phi^3 \right) .
\end{equation}
Thus, the third quantization formalism provides us with a system of interacting universes.
For instance, the propagator (\ref{A4}) would physically represent the bifurcation of
one universe (see, Fig. \ref{bifurcation}).

We can also define a wave function for the whole multiverse, i.e. $\Psi[\phi(X^i), X^0]$,
where $X^0$ is the field chosen as a time variable and then, $i = 1, ..., D-1$. It is given
by the solution of the third quantized Schr\"{o}dinger equation,
\begin{equation}\label{A7}
\mathrm{H} |\Psi \rangle = i \hbar \frac{\partial}{\partial X^0} |\Psi\rangle ,
\end{equation}
where $\mathrm{H}$ is the third quantized Hamiltonian (not to be confused with the Wheeler-De Witt operator (\ref{A2})). This is constructed from the action (\ref{A3}) with the momentum conjugate to the wave function of the universe defined by
\begin{equation}
P_\phi \equiv \frac{\delta \mathcal{L}}{\delta \dot{\phi}} = \sqrt{-G} \, G^{0\nu} \nabla_\nu \phi ,
\end{equation}
where, $\dot{\phi} \equiv \frac{\partial \phi}{\partial X^0} $ (in the general covariant
form, $P^\mu_\phi \equiv \frac{\delta ^3\mathcal{L}}{\delta \nabla_\mu \phi} =
\sqrt{-G} \, G^{\mu\nu} \nabla_\nu \phi $). The meaning of the wave function of the multiverse
given by Eq. (\ref{A7}) can then be understood as follows: let, $\{ |n\rangle\}_{|n=1, 2, ...}$,
be an orthonormal basis or number states, then
\begin{equation}
|\Psi\rangle = \sum_n \Psi(X^0) |n\rangle ,
\end{equation}
where, $\Psi(X^0)  \equiv \Psi(X^0, \phi(X^i))$, can be seen as the probability of $n$
universes at \emph{time} $X^0$ or the probability of $n$ universes with field values
$X^0$ \cite{Strominger90}.

This formalism for one-dimensional universes can be adapted for 3+1 dimensional
universes \cite{Strominger90}. The formalism has to be changed to account for the gauge
symmetries which are not present in the one-dimensional case. However, some of the concepts
of the one-dimensional case can be applied too to the case of more realistic universes.
Thus, from the point of view of the third quantization scheme, our single universe is just
one universe within the whole many-universe system. Nevertheless, in some approximation
we have to recover the single universe we inhabit. This consists of a large region of the
space-time of order of our Hubble length called a parent universe. Nevertheless,
within a single parent universe quantum fluctuations of the metric can also be considered.
Let us notice that the gravitational action depends on the length scale and at Planck length,
such fluctuations become of the same order of the metric \cite{Wheeler57}. Thus,
tiny regions of the space-time of order of the Planck scale can virtually tunnel out of their
parent space-time. These are called baby universes.

The parent universe is influenced by such quantum fluctuations of the gravitational vacuum,
for instance these can be seen as the source of a non-zero vacuum energy or cosmological constant,
$\Lambda$. In the third quantization scheme, therefore, we may say that a parent universe
propagates then in a plasma of baby universes. It can be depicted as follows:
let, $\phi_p$ and $\phi_b$, be the wave functions of a parent and a baby universe,
respectively. The second quantized action for each universe is given by Eq. (\ref{A1})
with masses $m_p$ and $m_b$, respectively. If we consider that the difference between the mass
scale of a parent and a baby universes is large, i.e. $m_p>>m_b$, then, quantum transitions
from baby to parent universes are exponentially suppressed.
However, Feynman diagrams like those represented in Fig. \ref{bifurcation} can still be considered.
The third quantized action which represents then a single parent universe propagating in a plasma
of baby universes is given by,
\begin{widetext}
\begin{equation}
S[\phi] = \frac{1}{2} \int d X \left( - (\nabla\phi_p)^2 + m_p^2 \phi_p^2 - (\nabla\phi_b)^2 + m_b^2 \phi_b^2 + \kappa \phi_p^2 \phi_b + \frac{\lambda}{3} \phi_b^3 \right) .
\end{equation}
\end{widetext}
Baby universes are not observable by themselves but their effects
may be observable, so it seems to be better to consider an hybrid
action. To do this we deal with $\phi_p$ as a wave function in the
second quantized formalism and with $\hat{\phi}_b$ as an operator
in the third quantized formalism. Then we can interpret baby
universes as tiny particles deployed in the parent space-time. The
result is an effective second quantized action for the parent
universe in which a potential term appears accounting for the
effects of the plasma of baby universes. In such a scheme, it can
be assumed an effective interaction action of the form,
\begin{equation}
S_I = \int d\tau \mathcal{N} \sum_i \mathcal{L}_i(\tau, \vec{x}) \phi_b^i ,
\end{equation}
where the index $i$ labels the different modes of the baby universe field
(i.e. it labels different species of baby universes), and $\mathcal{L}_i(\tau, \vec{x})$
is the insertion operator at the nucleation event. It defines the space-time points of the parent
universe in which the baby universes effectively nucleate.

\subsection{Homogeneous and isotropic universes}

Let us now sketch the third quantization formalism in a more
realistic model of the universe. Let us consider then the case of
an homogenous and isotropic space time with some matter fields
defined upon it. For the sake of simplicity let us assume just one
scalar field, $\varphi$ (the generalization to account for more
matter fields can easily be done). In that case, the second
quantized action can be written as \cite{Kiefer07}
\begin{equation}\label{A12}
S = \int dt [\frac{1}{2 \mathcal{N}} \mathcal{G}_{AB} (q)
\dot{q}^{A} \dot{q}^{B} - \mathcal{N} \mathcal{U}(q) ] ,
\end{equation}
where, $\mathcal{N}$ is again the lapse function and, $q^A = \{a(t), \varphi \}$, being $a(t)$
the scale factor. The minisupermetric is given then by,
\begin{equation}\label{A13}
\mathcal{G}_{AB} (a) = \left( \begin{array}{ccc} -\frac{3}{\sigma^2} a & 0  \\ 0 & a^3
\end{array} \right) ,
\end{equation}
with, $\sigma^2 = 4 \pi G = \frac{4 \pi}{m_p^2}$, being $G$ and $m_p$ the gravitational
constant and the Planck mass, respectively. The potential term in Eq. (\ref{A12}) is given by,
\begin{equation}
\mathcal{U}(q) = - \frac{1}{2\sigma^2} \kappa a + \frac{1}{2\sigma^2} \Lambda a^3 + a^3 V(\varphi) ,
\end{equation}
where, $\kappa = -1, 0, 1$, stands for hyperbolic, flat and closed spatial sections of the
space-time, respectively, $\Lambda$ is the cosmological constant and $V(\varphi)$ is the
potential of the matter field $\varphi$. Thus, for homogeneous and isotropic models it is
clear that the scale factor can play the role of a time variable. This can also be seen
from the signature of the minisupermetric (\ref{A13}). Therefore, the third quantization
procedure can formally be applied in a similar way to the case of one-dimensional universes.
Then the canonical momenta and Hamiltonian, respectively, are given by
\begin{eqnarray}\label{A15}
p_{A} & \equiv & \frac{\partial L}{\partial \dot{q}_{A}} =
\frac{1}{\mathcal{N}} \mathcal{G}_{AB} \dot{q}^{B}  , \\ \label{A16} H & = & \mathcal{N} [\frac{1}{2} \mathcal{G}^{AB} p_{A}
p_{B} + \mathcal{U}(q)] = \mathcal{N} \mathcal{H}.
\end{eqnarray}
The Hamiltonian constraint, $\frac{\delta H}{\delta \mathcal{N}} = 0$, turns out to be
$\hat{\mathcal{H}} \phi = 0$, where $\phi \equiv\phi(a, \varphi)$ is the wave function
of the universe. Finally, the third quantized action leading to the Wheeler-De Witt equation reads,
\begin{equation}
^3S = \frac{1}{2} \int da \left( \frac{1}{2} \mathcal{G}^{AB} \nabla_A \phi \nabla_B \phi + \mathcal{U} \phi^2 \right)
\end{equation}

As a particular example, let us consider a fluid with equation of state given by,
$p = w \rho$, being $w$ a constant parameter, and $p$ and $\rho$ the pressure and
the energy density of the fluid, respectively. These are given by,
\begin{eqnarray}\label{A17}
\rho &=& \frac{1}{2\mathcal{N}^2} \dot{\varphi} ^{2} + V(\varphi) , \\
p &=& \frac{1}{2\mathcal{N}^2} \dot{\varphi} ^{2} - V(\varphi) .
\end{eqnarray}
On the other hand, for an homogeneous and isotropic space-time the equation of conservation
of the cosmic energy reads,
\begin{equation}
d\rho = - 3 (p + \rho) \frac{da}{a} .
\end{equation}
It can be easily solved with the help of the equation of state, $p = w \rho$. The energy density turns
out to be then, $\rho(a) = c_0 a^{-3 (1+w)}$. Therefore, combining Eqs. (\ref{A17}) and (\ref{A16})
the Hamiltonian constraint turns out to be,
\begin{equation}\label{Friedmann}
p^2_a - \frac{\Lambda}{3} a^4 + \kappa  a^2 - \rho_0 a^{-3 w + 1} = 0 ,
\end{equation}
which is the Friedmann equation, being $p_a \equiv a \dot{a}$. Under canonical quantization, $\hat{p}_a \rightarrow -i \hbar \frac{\partial}{\partial a}$, and the Wheeler-De Witt equation can be written as,
\begin{equation}\label{WDW}
\ddot{\phi} + \omega^2(a) \phi = 0 ,
\end{equation}
where, $\phi\equiv \phi(a)$, is the wave function of the universe. Eq. (\ref{WDW}) can be seen as the classical equation of motion of a harmonic oscillator. It can be derived from the following third quantized action,
\begin{equation}\label{3rd action}
^3S = \frac{1}{2} \int da \left( \dot{\phi}^2 - \omega^2(a) \phi^2 \right) ,
\end{equation}
which formally corresponds to the action of a harmonic oscillator, where the scale factor $a(t)$ formally plays the role of time. The \emph{time-dependent} frequency, $\omega(a)$, is given by
\begin{equation}\label{frequency}
\omega(a) = \frac{1}{\hbar} \sqrt{\frac{\Lambda c^2}{3} a^4 - \kappa c^2  a^2 + \rho_0 a^{-3 w + 1}}  \, ,
\end{equation}
where the velocity of light has been introduced for completeness.

In the general case the computations are actually more complicated. However, the Wheeler-De Witt
equation can be seen like a wave equation and therefore, some of the developments used in
quantum optics might also be applied to the third quantized formalism of the multiverse.
For instance, let us make the change, $\alpha\equiv \ln a$. Then, the Wheeler-De Witt
equation can be written as \cite{Kiefer07}
\begin{equation}\label{A21}
- \frac{\partial^2 \phi}{\partial \alpha^2} + \frac{\partial^2 \phi}{\partial \varphi^2} = \mathcal{J}(\alpha, \varphi) \phi .
\end{equation}
This can be seen as the wave equation for a radiation field \emph{propagating in the void}
(see, Fig. \ref{comparative}), at the speed $c=1$, where the field is now the wave function of the universe,
$\phi\equiv\phi(\alpha, \varphi)$. The \emph{source term} \cite{Scully97} in Eq. (\ref{A21}),
$\mathcal{J}(\alpha, \varphi)$, is given by
\begin{equation}\label{A22}
\mathcal{J}(\alpha, \varphi) = \frac{e^{4\alpha}}{\hbar^2} \left( \left(\frac{\Lambda}{3} + V(\varphi) \right) e^{2\alpha} - \kappa \right).
\end{equation}
Now, let us expand the modes of the \emph{free} wave function of the universe ($\mathcal{J}=0$,
in Eq. (\ref{A21})) as,
\begin{equation}\label{A23}
\phi(\alpha,\varphi) = \sum_k b_k  e^{-i \omega_k \alpha + i k \varphi} + \rm{c.c.} .
\end{equation}
Then, the third quantization procedure eventually consists of promoting the wave function of the
universe into an operator, $\hat{\phi}$. Therefore, let $\hat{b}_k^\dag$ and $\hat{b}_k$ be, respectively,
the creation and annihilation operators of universes in the mode $k$, i.e.
\begin{eqnarray}\label{A24}
\hat{b}_k & = & \sqrt{\frac{\omega_k}{2 \hbar}} \left( \hat{\phi} + \frac{i}{\omega_k}  \hat{p}_\phi \right) , \\  \label{A25}
\hat{b}_k^\dag & = & \sqrt{\frac{\omega_k}{2 \hbar}} \left( \hat{\phi} - \frac{i}{\omega_k}  \hat{p}_\phi \right) ,
\end{eqnarray}
Thus, the system described by Eq. (\ref{A21}) can formally be viewed as the radiation
emitted by a classical current distribution given by $\mathcal{J}$. Following the analogy,
the Hamiltonian that would describe the interaction between the field $\phi$ and the current
$\mathcal{J}$ is given by \cite{Scully97},
\begin{equation}
\mathcal{V}(\alpha) = \int_{-\infty}^\infty d\varphi \,\, \mathcal{J}(\alpha, \varphi) \phi(\alpha, \varphi) .
\end{equation}
In the interaction picture, therefore, the state of the multiverse
evolves (as the scale factor is growing up) into a coherent state
defined by,
\begin{equation}
\frac{\partial }{\partial \alpha} |\Psi_I(\alpha)\rangle = e^{- \frac{i}{\hbar} \int^\alpha \mathcal{V}(\alpha') d\alpha'    } |\Psi_I(0)\rangle ,
\end{equation}
where \cite{Scully97},
\begin{equation}
e^{- \frac{i}{\hbar} \int^\alpha \mathcal{V}(\alpha') d\alpha' }  = \prod_k e^{\beta_k \hat{b}_k^\dag - \beta_k^* \hat{b}_k} ,
\end{equation}
with
\begin{equation}
\beta_k(\alpha) = \int^\alpha d\alpha' \int_{-\infty}^\infty d\varphi \,\, \mathcal{J}(\alpha, \varphi) e^{i\omega_k \alpha - i k \varphi} .
\end{equation}
Let us take as the boundary condition for the \emph{initial} state of the multiverse that for a
zero value of the scale factor there is no space, time and matter (i.e. there is 'nothing').
Analogously, this is what Strominger calls the \emph{void} \cite{Strominger90}, i.e. $|0\rangle$.
Then, the state of the multiverse described by Eq. (\ref{A21}) would evolve as the scale factor is
growing up into a coherent state given by,
\begin{equation}
|\Psi(\alpha)\rangle = \prod_k e^{\beta_k \hat{b}_k^\dag - \beta_k^* \hat{b}_k} |0\rangle_k .
\end{equation}
This is an example of how far can be formally taken the analogy between the electromagnetic field
and the third quantization formalism of the quantum multiverse, at least in the case of cosmological
models of high symmetry.

\section{The wave function of the multiverse}

Let us consider a quantum multiverse made up of homogeneous and isotropic universes filled up with a perfect fluid, for which the Friedmann equation is given by Eq. (\ref{Friedmann}). Following the third quantization procedure, the third quantized action for such a many-universe system is given by Eq. (\ref{3rd action}). It formally corresponds to the classical action of a harmonic oscillator, where it is the scale factor which plays the role of time. Then, the quantum state of the multiverse, $|\Psi \rangle$, can be defined as the solution of the third quantized Schr\"{o}dinger equation,
\begin{equation}\label{Schrodinger third quantized}
\mathbf{\mathrm{H}} |\Psi \rangle = i \hbar \frac{\partial}{\partial a} |\Psi \rangle ,
\end{equation}
where
\begin{equation}\label{Hamiltonian 3-quantized}
\mathbf{\mathrm{H}} = \frac{1}{2} p_\phi^2 + \frac{\omega^2(a)}{2}
\phi^2 ,
\end{equation}
$p_\phi$ being the momentum conjugate to the wave function of a single universe, $\phi$. The frequency $\omega(a)$ is given by Eq. (\ref{frequency})

Different disconnected regions of the whole space-time may undergo different expansion (or contraction) rates because they can be filled up with different energy-matter contents and have different spatial geometries, both features being incorporated in the value of the frequency $\omega(a)$. Therefore, given an orthonormal basis of solutions of Eq. (\ref{Schrodinger third quantized}), $\Psi_{N}^{\omega}(\phi)$, the general quantum state of a multiverse made up of disconnected regions, each being dominated by a particular kind of energy-matter content, can be expressed in terms of a linear combination of product states of the form
\begin{equation}\label{stateMultiverse}
\Psi_m = \Psi_{N_1}^{\omega_1}(\phi_1)  \Psi_{N_2}^{\omega_2}(\phi_2) \cdots  \Psi_{N_n}^{\omega_n}(\phi_n) ,
\end{equation}
where $N_i$ is the number of universes of type $i$, represented by a second quantized wave function $\phi_i$, which are filled up with an energy-matter content described by the frequency $\omega_i(a)$. Let us notice that the quantum state given by Eq. (\ref{stateMultiverse}) is a well-defined state for closed geometries and open geometries.

The frequency $\omega(a)$ of the third quantized harmonic oscillator which would represent the quantum state of the multiverse, contains not just the information of the second quantized theory but also that of the classical one. Using the Friedmann equation we can show that the frequency determines the expansion rate of a single universe. For instance, let us first consider the case of a flat de Sitter universe with an exponential expansion. The solutions of Eq. (\ref{Friedmann}) are then given by, $a(t) = a_0 e^{\pm \omega_0 t}$, with $\omega_0 = \sqrt{\frac{c^2 \Lambda}{3}}$, where the positive branch corresponds to the exponential expansion of a flat de Sitter space-time. Taking into account the negative branch, the solution for a closed de Sitter space-time can be obtained \cite{Vilenkin82}, i.e. $a(t) = a_0 \cosh \omega_0 t$. Furthermore, in conformal time, $\eta = \int \frac{d t}{a(t)}$, the scale factor turns out to be, $a(\eta) = \frac{a_0}{\cos\eta}$, which can be seen as the two-dimensional metric of a cosmological warp drive \cite{GonzalezDiaz09}. Next we consider the anti-De Sitter space-time for which $\Lambda < 0$. Analogously to the former case, the solution $a(t) = a_0 \cos \omega_0 t$ can be obtained, which in conformal time turns out to be, $a(\eta) = \frac{a_0}{\cosh \omega_0 \eta}$, and as a final example let us note that the Tolman-Hawking closed baby universe used in \cite{GonzalezDiaz09} corresponds to the values, $\kappa = 1$ and $\rho = 0$, in Eq. (\ref{Friedmann}). In a similar way we can depict other kinds of baby and parent universes in the third quantized theory through the value of the frequency.

Quantum mechanically, in the case of a multiverse made up of flat universes filled up with a radiation like fluid, for which $w = \frac{1}{3}$, the frequency $\omega(a)$ turns out to be a constant given by $\omega = \frac{\omega_0}{\hbar}$, and the state of the multiverse is represented then by the state of a harmonic oscillator with constant frequency which is, of course, very well-known. In general, for universes with fluids for which $w\neq \frac{1}{3}$, the quantum state of the multiverse can be expressed in terms of harmonic oscillators with a \emph{time}-dependent frequency, which have been studied in the past \cite{Lewis69, Pedrosa87}, and to which a renewed attention has been recently paid \cite{Vergel09}. Using the method of invariants developed by Lewis and Riesenfeld \cite{Lewis69}, the solutions of Eq. (\ref{Schrodinger third quantized}) can be expressed in terms of the eigenstates of an auxiliary operator, $I$, which is invariant under the evolution (i.e. scale factor invariant) given by the Hamiltonian, $\mathrm{H}$, i.e. $\frac{d I}{d a} = i \hbar \frac{\partial I}{\partial a} - [\mathrm{H}, I] = 0$, which implies that, $i \hbar \frac{\partial}{\partial a} I |\rangle = \mathrm{H} I |\rangle$, with $| \rangle$ being a Schr\"{o}dinger state vector. Therefore, the action of an invariant operator $I$ on a Schr\"{o}dinger state vector produces another solution of the Schr\"{o}dinger equation. Furthermore, if the invariant operator does not involve differentiation under the scale factor, relative phases for the eigenstates of $I$ can be found in such a way that such states themselves satisfy the Schr\"{o}dinger equation  (\ref{Schrodinger third quantized}), giving us an orthonormal basis for the space of solutions. Thus, let us assume following \cite{Lewis69}  the existence of a Hermitian invariant operator, $I$, with the homogeneous, quadratic form, $I = \frac{1}{2} [ \alpha \phi^2 + \beta p_\phi^2 + \gamma \{\phi, p_\phi \}_+ ]$. For such an operator to be invariant a set of equations has to be satisfied by the coefficients $\alpha$, $\beta$ and $\gamma$. Solving those equations, the invariant operator can be written as, $I = \frac{1}{2} [ (\frac{1}{R^2}) \phi^2 + ( R p_\phi - \dot{R} \phi )^2 ]$, where the function $R \equiv R(a)$ obeys the auxiliary equation
\begin{equation}\label{Requation}
\ddot{R} + \omega^2(a) R - \frac{1}{R^3} = 0 .
\end{equation}
Eq. (\ref{Requation}) can also be solved in terms of the solutions of the Wheeler-De Witt equation, $\ddot{\phi} + \omega^2(a) \phi = 0$. Let $\phi_1$ and $\phi_2$ be two independent solutions of the latter equation, then
\begin{equation}\label{R}
R(a) \equiv \sqrt{\phi_1^2(a) + \phi_2^2(a) } ,
\end{equation}
is a solution of Eq. (\ref{Requation}). Moreover, the invariant operator $I$ can be expressed in terms of creation and annihilation operators, such as
\begin{eqnarray}\label{creationAnnihilation1}
b(a) & = & \sqrt{\frac{1}{2 \hbar}} \left(\frac{\phi}{R} + i (R  p_\phi - \dot{R} \phi ) \right) , \\  \label{creationAnnihilation2}
b^\dag(a) & = & \sqrt{\frac{1}{2 \hbar}} \left(\frac{\phi}{R} - i (R p_\phi - \dot{R} \phi ) \right) ,
\end{eqnarray}
which gives, $I = \hbar (b^\dag b + \frac{1}{2})$. For a flat space-time we can obtain analytical solutions of Eq. (\ref{Requation}) and $R$ is given by Eq. (\ref{R}) with
\begin{equation}\label{fi1}
\phi_1(a) = \sqrt{\frac{\pi a}{2 q}} J_{\frac{1}{2q}}\left(\frac{\omega_0 a^q }{q \hbar} \right) , \;\; \phi_2(a) = \sqrt{\frac{\pi a}{2 q}}
Y_{\frac{1}{2q}}\left( \frac{\omega_0 a^q }{q \hbar}\right) ,
\end{equation}
where $J_n(x)$ and $Y_n(x)$ are Bessel functions of first and second kind, respectively, of order $n$. For other spatial geometries and energy-matter contents, the Wheeler-De Witt equation (\ref{WDW}) may not be analytically solvable. However, it does not introduce any conceptual drawback as far as the existence of the solutions of Eq. (\ref{WDW}) is guaranteed, and they may be obtained analytically or numerically. Nevertheless, analyticity makes clearer the computations and for that reason, we will mainly consider the case of flat universes for which such solutions are given by Eq. (\ref{fi1}). In any case the scale factor dependent creation and annihilation operators given by Eqs. (\ref{creationAnnihilation1}) and (\ref{creationAnnihilation2}) give the expected results when they are applied on the eigenstates of the operator $I$, $| N, a \rangle$, i.e.
\begin{eqnarray} \label{numberstates1}
b(a) | N, a \rangle &=& \sqrt{N} | N - 1, a \rangle , \\ \label{numberstates2} b^\dag(a) | N , a \rangle &=& \sqrt{N+1} | N +1, a \rangle , \\ \label{numberstates3}
b^\dag(a) b(a) | N ,a \rangle &=& N | N ,a \rangle ,
\end{eqnarray}
and, $I | N, a \rangle = \hbar (N + \frac{1}{2}) | N, a \rangle$. Therefore, the eigenvalues are invariant under scale factor transformations. Furthermore, the scale factor dependent eigenstates, $ | N, a\rangle$, can be linked to the eigenstates of a harmonic oscillator with constant frequency (let us say, $\omega = 1$). In the Schr\"{o}dinger representation, both wave functions are related through the following unitary relation,
\begin{equation}\label{eigenfunctions}
\Psi_N (\phi, a) = \frac{1}{\sqrt{R(a)}} \,
U^\dag \, \tilde{\Psi}_N(\varphi)\mid_{\varphi=\frac{\phi}{R(a)}} ,
\end{equation}
where, $\Psi_N (\phi, a) \equiv \langle \phi | N, a \rangle$, and $\tilde{\Psi}_N(\varphi)$ are the eigenfunctions of the Hamiltonian of a harmonic oscillator with constant frequency, i.e. $\mathbf{\mathrm{H}}_0 \tilde{\Psi}_N(\varphi) = \hbar (N + \frac{1}{2}) \tilde{\Psi}_N(\varphi)$, with $\mathbf{\mathrm{H}}_0 = \frac{1}{2} p_\varphi^2 + \frac{1}{2} \varphi^2 $. The unitary operator, $U\equiv U(\phi, a)$, is given by
\begin{equation}
U (\phi, a) = e^{-\frac{i}{2 \hbar} \frac{\dot{R}}{R} \phi^2} ,
\end{equation}
with, $R \equiv R(a)$, defined by Eq. (\ref{R}). Then, the most general solution $| \Psi \rangle$ of the Schr\"{o}dinger equation (\ref{Schrodinger third quantized}), can be written as \cite{Pedrosa87}
\begin{equation}\label{state_multiverse}
|\Psi , a \rangle = \sum_N C_N e^{i \alpha_N(a)} \Psi_N(\phi, a) |N , a \rangle ,
\end{equation}
where the relative phases, $\alpha_N(a)$, are given by
\begin{equation}\label{phases}
\alpha_N(a) = - (N + \frac{1}{2}) \int_0^a \frac{da'}{R^2(a')} ,
\end{equation}
which has a well defined value because the zeros of the Bessel functions $J_n(x)$ and $Y_n(x)$ do not coincide, and therefore $R$ is not degenerate for any value of the scale factor. The number eigenstates $| N, a \rangle$ of the invariant operator $I$ form an orthonormal basis for the space of solutions of the Schr\"{o}dinger equation, and the eigenfunctions $\Psi_N(\phi, a)$ in Eq. (\ref{state_multiverse}) can thus be interpreted as the probability amplitude for $N$ universes with a value of the scale factor $a$, for which their second quantized wave functions are given by $\phi$. If we consider the quantum state of a set of universes with different values of the scale factor, the probability amplitudes would be given by the product of harmonic wave functions (\ref{state_multiverse}), for each value of the scale factor. However, it seems to be more appropriate to consider probability amplitudes regardless of the value of the scale factor because, as it has been already noticed, the number of universes within the multiverse is a scale factor invariant quantity. In that case, we integrate over the scale factor to trace it out. On the other hand, since $\phi$ has a well-defined semi-classical regime \cite{GonzalezDiaz07}, the quantum state of the multiverse given by Eq. (\ref{state_multiverse}) can represent both a multiverse of homogeneous and isotropic parent universes and a space-time foam made up of baby universes. Moreover, let us notice that unlike in the second quantized theory, in which a quantum state of the universe cannot in general be defined \cite{Hartle94}, the third quantized state of the multiverse given by Eq. (\ref{state_multiverse}) is indeed well-defined because there are no space-time singularities in the quantum multiverse.

However, the orthonormal eigenstates $|N, a\rangle$, are not eigenstates of the Hamiltonian given by Eq. (\ref{Hamiltonian 3-quantized}) for all values of the scale factor (except in the case $w=\frac{1}{3}$), and therefore they are not stationary solutions for the quantum state of the multiverse. In fact, in terms of the creation and annihilation operators given by Eqs. (\ref{creationAnnihilation1}) and (\ref{creationAnnihilation2}), the Hamiltonian (\ref{Hamiltonian 3-quantized}) can be written as
\begin{equation}\label{Hamiltonian2}
H = \hbar \left[ \beta_- b^2 + \beta_+{b^\dag}^2 + \beta_0 \left( b^\dag b + \frac{1}{2} \right) \right] ,
\end{equation}
where,
\begin{eqnarray}\label{beta+}
\beta_+^* = \beta_- &=& \frac{1}{4} \left\{ \left( \dot{R} - \frac{i}{R} \right)^2 + \omega^2 R^2  \right\} , \\ \label{beta0}
\beta_0 &=& \frac{1}{2} \left( \dot{R}^2 + \frac{1}{R^2} + \omega^2 R^2 \right) .
\end{eqnarray}
The quadratic terms in $b^\dag$ and $b$ in the Hamiltonian given
by Eq. (\ref{Hamiltonian2}) make the quantum state of the
multiverse to evolve into a squeezed state \cite{Walls83}. The squeezing effect is greater for the quantum state of a multiverse
made up of accelerated universes \cite{RoblesPerez08}. It might
actually be related to the quantum nature of accelerated universes \cite{RoblesPerez08, GonzalezDiaz09}, in the sense that squeezed states are usually considered
to be sharp quantum states with no classical analog \cite{Walls83, Reid86}.
However, the squeezing effect vanishes as the dominant fluid of
the single universes that form up the multiverse becomes more and
more similar to a radiation fluid when $w\rightarrow \frac{1}{3}$.
Then, the state of the multiverse evolves into a set of
conventional coherent states, which are usually taken to be the
most classically possible states in the quantum theory. That can
also be considered in the case of the multiverse, because for $w
=\frac{1}{3}$ the wave function of a single universe turns out to
be a plain wave whose argument is the classical action, i.e.
$\phi_{rad}(a) = e^{\pm \frac{i}{\hbar} S_c(a)}$, with $S_c(a) =
\omega_0 a$. Then, the Hamilton-Jacobi equation and the classical
equation of motion for the scale factor are satisfied \cite{Halliwell87}, being the
distribution of momenta a delta function centered at the classical
momentum, i.e. $\phi_{rad}(p) = \delta(p-p_c)$, with $p_c =
\omega_0$ the classical momentum. In that sense, $\phi_{rad}$
represents a classical universe and coherent states can
associated with a multiverse consisting of classical universes (see
also Ref. \cite{RoblesPerez08}).

Therefore, for a general value of the scale factor and $w \neq \frac{1}{3}$, the state of the multiverse is represented by a squeezed state. However, taking into account the asymptotic expansions of Bessel functions for large arguments in the case of parent universes with a large value of the scale factor, in Eqs. (\ref{R}) and (\ref{fi1}), it can be shown that $R(a) \approx \omega^{-\frac{1}{2}}(a)$, being $\omega(a) = \omega_0 a^{q-1}$. Thus, the coefficients $\beta_\pm$ and $\beta_0$ in the Hamiltonian (\ref{Hamiltonian2}) can be approximated to
\begin{eqnarray}
\beta_+^* = \beta_- & \approx & \frac{i (q-1)}{4 a} \rightarrow 0 , \\
\beta_0 & \rightarrow & \omega(a) ,
\end{eqnarray}
so that the Hamiltonian (\ref{Hamiltonian2}) turns out to be the usual Hamiltonian for a harmonic oscillator in terms of the number operator, $N \equiv b^\dag b$, with a scale factor dependent frequency given by $\omega(a)$. Then, the quantum correlations between number states disappear and they turn out to be an appropriate representation for the quantum state of the multiverse. In particular, at large values of the scale factor (i.e.; $\frac{a}{\hbar} \gg 1$), the single universe approximation for which $\langle N \rangle = 1$, can be considered. On the other hand, taking into account the expansion of Bessel functions for small arguments in the case of virtual baby universes with a small value of the scale factor (typically of Planck length)  in Eqs. (\ref{R}) and (\ref{fi1}), then, $\phi_1 \rightarrow 0$ and $\phi_2 \rightarrow \nu_0^{-\frac{1}{2}}$, and therefore, $R \rightarrow \nu_0^{-\frac{1}{2}}$ and $\dot{R}\rightarrow 0$ for $q>0$. Thus, from Eqs. (\ref{beta+}) and (\ref{beta0}), it follows that the coefficients $\beta_\pm$ and $\beta_0$ can be approximated as
\begin{eqnarray}
\beta_+^* = \beta_-  & \rightarrow & -\frac{\nu_0}{4} , \\
\beta_0 & \rightarrow & \frac{\nu_0}{2} ,
\end{eqnarray}
where $\nu_0$ is a constant parameter given by,
\begin{equation}\label{nu_0}
\nu_0 = \frac{\pi (2 q)^{\frac{q-1}{q}} }{\Gamma^2(\frac{1}{2 q})} (\frac{\omega_0}{\hbar})^{\frac{1}{q}} ,
\end{equation}
if $q \neq 1$. For $q = 1$, the value of $\nu_0$ actually coincides  with $\hbar^{-1}\omega_0$, but in that case $\beta_\pm = 0$ in the Hamiltonian given by Eq. (\ref{Hamiltonian2}) and no squeezing effect is present, as has been already shown. However, if we assume that baby universes are instead vacuum dominated universes or closed baby universes, for which $w= -1$ and $w=-\frac{1}{3}$, respectively, then the non-vanishing value $\beta_+$ and $\beta_-$ in Eq. (\ref{Hamiltonian2}) when the scale factor decreases implies that high order correlations between number states are going to play an extremely important role in the space-time foam of baby universes at the Planck length. It also means therefore that the number state representation is not an appropriate representation of the quantum gravitational vacuum. Instead, it would be more properly represented by a squeezed state, which might be even experimentally observable because such a gravitational state would induce a loss of quantum coherence in the vacuum state of the matter field sector \cite{GonzalezDiaz92}. In fact, from the Hamiltonian (\ref{Hamiltonian2}) one can clearly infer that baby universes should be continuously created and annihilated in pairs, at least under the assumptions made in the present model. That appears to enhance the result that euclidean doubly connected wormholes copiously nucleate, together perhaps with simply connected wormholes. This implication is relevant for the debate on the loss of quantum coherence in the matter field sector due to the gravitational vacuum (see Refs. \cite{GonzalezDiaz92} and \cite{Coleman88}) because it suggests that the loss of quantum coherence of matter fields might be unavoidable.

On the other hand, the third quantized vacuum energy, $E_3 = \frac{\hbar \nu_0}{2}$, retains a marginal non zero value even when the scale factor decreases to $a=0$, i.e -- when the universes are annihilated. This is a well-known feature of the quantum vacuum state with no classical analog, applied in the present case to the third quantized vacuum state of gravity.

Finally, let us briefly discuss the question on how and when the universes are quantum cosmologically created. In the third quantization formalism, an interaction scheme can be used to describe the space-time foam of a parent universe. As it was shown in Sec. II, an insertion operator is then introduced to define the space-time points of the parent universe in which the creation and annihilation operators of baby universes are located. Thus, the interaction Hamiltonian can in general be written as, $H_{int} = \sum_i f_i(t,\vec{r}) g(b^\dag_i, b_i)$, where the index $i$ accounts for the different species of baby universes being considered in the space-time foam.

However, in the case of a multiverse made up of parent universes, which cannot be placed in a common space-time, the above question becomes misleading. Let us notice that space and time are properties with a well-defined meaning only within a single parent universe, and they are not therefore well-defined \emph{outside} the universe. So the Hilbert space corresponding to a multiverse made up of parent universes cannot  be described in terms of space-time coordinates but, instead, of statistical ones. However, correlated pairs of universes can still be considered because entanglement is a quantum property directly stemming from the superposition principle, and therefore, a pair of parent universes can be created at points of such a general statistical Hilbert space.

\section{The density matrix}

\subsection{The second quantized wave function representation}

The most general quantum state of the multiverse is not represented by a pure state defined by a third quantized wave function, but rather by a mixed state described in terms of a density matrix operator, $\rho$, for which the explicit scale factor dependence is given in the third quantization formalism by the Schr\"{o}dinger like equation, $\frac{\partial }{\partial a} \langle \rho \rangle = \frac{i}{\hbar} \langle [H, \rho] \rangle$. Therefore, let us consider the quantum state of a multiverse made up of flat space-times in the representation defined by the value of the second quantized wave function $\phi$, i.e. the \emph{coordinate} representation. Then, taking into account the expression of the Hamiltonian given by Eq. (\ref{Hamiltonian 3-quantized}), the Schr\"{o}dinger equation for the density matrix in such a representation can be written as,
\begin{widetext}
\begin{equation}\label{schrodinger3}
\frac{\partial \rho(\phi, \phi', a)}{\partial a} =  \frac{i}{\hbar} \left(  -\frac{\hbar^2}{2} \left( \frac{\partial^2}{\partial \phi^2} - \frac{\partial^2}{\partial \phi'^2} \right) + \frac{\omega^2(a)}{2}  (\phi^2 - \phi'^2)   \right)  \rho(\phi, \phi', a).
\end{equation}
\end{widetext}
This can be solved by using the following Gaussian ansatz,
\begin{equation}\label{ansatz}
 \rho(\phi, \phi', a) =  e^{ - A(a) (\phi + \phi')^2 - i B(a) (\phi^2 - \phi'^2) - C(a) } .
\end{equation}
Inserting Eq. (\ref{ansatz}) into the Schr\"{o}dinger equation (\ref{schrodinger3}), we obtain the following equations for the coefficients,
\begin{eqnarray}\label{1}
\dot{A} &=& - 4 \hbar B A, \\ \label{2}
\dot{C} &=&  2 \hbar B , \\ \label{3}
\dot{B} &=& - 2 \hbar B^2 - \frac{\omega^2}{2 \hbar} .
\end{eqnarray}
The last of these equations can actually be transformed into the Wheeler-De Witt equation by making the following change: $x = e^{2 \hbar \int B da}$ (i.e. $B = \frac{1}{2 \hbar} \frac{\dot{x}}{x}$). Then, Eq. (\ref{3}) turns out to be $\ddot{x} + \omega^2 x = 0$, the solutions of which are given by Eq. (\ref{fi1}). The quantum state of the multiverse can therefore be expressed in terms of the two values of the second quantized wave functions of the universe, $\phi$ and $\phi'$ respectively, as
\begin{equation}\label{generaldensity}
\rho(\phi, \phi', a) = \frac{C_0}{\phi_0 } e^{-\frac{\pi C_0^2}{4 \phi_0^2} (\phi +\phi')^2 - \frac{i}{2 \hbar} \frac{\dot{\phi}_0}{\phi_0}(\phi^2 - \phi'^2) } ,
\end{equation}
where the constants of integration have been chosen to satisfy the normalization criteria, $\int d\phi |\rho(\phi,\phi,a)| = 1, \; \forall a$, and, $\phi_0 \equiv \phi_0(a) = A_0 \phi_1(a) + B_0 \phi_2(a)$, is the general solution of the Wheeler-De Witt equation. $A_0$ and $B_0$ are two constants to be determined by the boundary conditions. The trace of the density matrix given by Eq. (\ref{generaldensity}) shows no divergences even when $\phi_0 \rightarrow 0$, it vanishes for an infinite value of the scale factor, avoiding in this way cosmic singularities, and it also satisfies the usual boundary conditions imposed to the second quantized wave function of the universe when the scale factor degenerates as $a \rightarrow 0$. This can be checked by taking into account the limits of the Bessel functions in Eq. (\ref{fi1}) for a vanishing value of the scale factor, i.e.  $\phi_1(a) \rightarrow 0$ and $\phi_2(a) \rightarrow \nu_0^{-\frac{1}{2}}$. Thus, if $B_0=0$, the solution of the Wheeler-De Witt equation is given by $\phi_1(a)$, and the density matrix (\ref{generaldensity}) tends to zero as the scale factor vanishes, satisfying in this way a boundary condition which corresponds to the Hartle-Hawking's no-boundary condition \cite{Hartle83}. On the other hand, if $B_0 \neq 0$, then $\phi_0 \rightarrow B_0 \nu^{-\frac{1}{2}}$, and the density matrix (\ref{generaldensity}) still represents a small oscillation even when the scale factor decreases to $a=0$, satisfying then a boundary condition which corresponds to the Vilenkin's proposal of a tunneling boundary condition \cite{Vilenkin86}. The state of the multiverse is therefore well-defined by Eq. (\ref{generaldensity}), and the diagonal elements, $|\rho(\phi,\phi,a)|$, can be then interpreted as a probability distribution which, for a given value $\phi$, is peaked around the solution of the Wheeler-De Witt equation, $\phi_0(a)$, for which the distribution $|\rho(\phi,\phi,a)|$ is a maximum, i.e. $\partial_{\phi_0} \rho(\phi,a)_{|\phi_0 = \phi}=0$.

Let us recall that the state of the multiverse represented by the density matrix (\ref{generaldensity}) avoids cosmic singularities, at $a=0$ and $a=\infty$, because it gives a zero value of the probability amplitude both when the scale factor degenerates and when it grows up to infinity. For instance, let us consider the case of a phantom energy dominated universe in which a bip rip singularity appears in the future of our proper time, splitting the universe into two causally disconnected parts before and after the singularity. In the context of the third quantized multiverse, such a model is rather described by two universes, provided that their quantum states are well defined by Eq. (\ref{generaldensity}), which might be even entangled \cite{GonzalezDiaz07}, and therefore no singularities appear in such a multiverse. That is an example of the way in which the third quantized formalism can also account for topological changes in the space-time, like the one that would happen, for instance, in the transition from a matter dominated universe into a phantom dominated one (i.e. from one single universe to a pair of, may be entangled, universes).

Let us also notice that the quantum state represented by the density matrix (\ref{generaldensity}) is a non-factorizable state for any value of the scale factor. However, it cannot be inferred that the quantum state of the multiverse is then in a mixed state. It would only be so if the representation given by the second quantized wave function of a single universe is orthogonal. Nevertheless, the solutions of the Wheeler-De Witt equation are not in general orthogonal \cite{GonzalezDiaz07}, and therefore it is not at all clear in such a representation whether the multiverse stays in a pure or a mixed state. For instance, for large values of the scale factor the quantum state represented by the density matrix (\ref{generaldensity}) is still a non-factorizable state. However, it has been shown that in the same limit the number state representation is orthogonal and diagonalizes the Hamiltonian. Thence, the  quantum state of a single parent universe is described by a pure state defined by the value of the quantum number, $N=1$. The number state representation could be seen then as an orthogonal or \emph{decoherenced} coarse graining \cite{Hartle91}, given by a superposition of second quantized wave functions weighted along $\phi$ accordingly to the third quantized wave functions of a harmonic oscillator.

Finally, let us recall that the density matrix for the quantum state of the universe was already obtained in Ref. \cite{Page86}. Our consideration of the density matrix in this section can be just viewed then for the aim of completeness concerning the properties of the quantum state of the multiverse in the third quantization formalism.

\subsection{The squeezed number representation}

Let us consider now the number state representation defined by the relations given in Eqs. (\ref{numberstates1}-\ref{numberstates3}). It has been shown that in the case of parent universes with a large value of the scale factor, or equivalently in the semi-classical limit where $\frac{a}{\hbar} \rightarrow \infty$, such a number representation diagonalizes the Hamiltonian $H$, because $\beta_\pm \rightarrow 0$ in Eq. (\ref{Hamiltonian2}). In that case, the scale factor dependence of the density matrix, $\rho = \sum_{N,M} P_{NM} |N, a\rangle \langle M, a |$, is given by the Schr\"{o}dinger equation which implies a differential equation for the density matrix elements, i.e. $i \dot{P}_{NM} = \beta_0 (M-N) P_{NM}$, where the dot stands for the derivative with respect to the scale factor. The solutions are given then by,
\begin{equation}\label{Pnumber}
P_{NM}(a) = e^{-i(M-N) \int^a da' \beta_0(a')} P_{NM}(0) ,
\end{equation}
where $\beta_0(a)$ is again defined by Eq. (\ref{beta0}), with $\beta_0(a) \approx \omega(a) = \frac{\omega_0 a^{q-1}}{\hbar}$ for large values of the scale factor. Therefore, if the initial density matrix is diagonal in such a number representation, i.e. $P_{NM}(0) = \delta_{NM} P_N$, then the diagonal elements remain constant with respect to the scale factor. For instance, for the pure state given by $P_N = \delta_{1N}$, which would represent a large single universe, there are no off-diagonal elements in the representation of the density matrix as the scale factor continues to increase and, therefore, once a large universe can be considered, transitions to other number of universes are asymptotically suppressed along the subsequent expansion of such a single universe.

However, for small values of the scale factor in such a representation, non-diagonal elements appear in the Schr\"{o}dinger equation and the number states turn out to be squeezed states. Let us note that in terms of the creation and annihilation operators for modes of a unity value of the frequency, $b_0^\dag \equiv \frac{1}{\sqrt{2 \hbar}} (\phi - i p_\phi) $ and $b_0 \equiv \frac{1}{\sqrt{2 \hbar}} (\phi + i p_\phi)$, the creation and annihilation operators given by Eqs. (\ref{creationAnnihilation1}) and (\ref{creationAnnihilation2}) can be written as,
\begin{eqnarray}
b(a) &=& \mu_0 b_0 + \nu b_0^\dag , \\ b^\dag(a) &=& \mu_0^* b_0^\dag + \nu_0^* b_0 ,
\end{eqnarray}
where,
\begin{eqnarray}\label{mu0}
\mu_0 &=& \frac{1}{2} \left(  \frac{1}{R} + R - i \dot{R} \right) , \\ \label{nu0} \nu_0 &=& \frac{1}{2} \left( \frac{1}{R} - R - i \dot{R} \right),
\end{eqnarray}
with, $|\mu_0|^2 - |\nu_0|^2 = 1$, and $R \equiv R(a)$ defined by Eq. (\ref{R}). It means therefore that the number eigenstates of the invariant operator $I$ do not diagonalize the Hamiltonian of the multiverse, and the squeezing effect becomes then larger for small values of the scale factor. However, let us also notice that quantum effects like squeezing or entanglement crucially depend on the chosen representation \cite{Vedral06}. This can be easily shown in the realm of quantum optics with a simple example: one photon state in a given mode of constant frequency, namely $\omega_1$, which is of course a number eigenstate in the above representation, turns out to be a squeezed state in the representation of the number states given for another different frequency, say $\omega_2$, i.e. $b_1 = \mu b_2 + \nu b_2^\dag$, with $|\mu|^2 - |\nu|^2 = 1$, being $b_{1,2}$ and $b_{1,2}^\dag$ the annihilation and creation operators of each mode, respectively. In quantum optics the appropriate representation is derived from the experimental setting, e.g. if the experimental devices of the setting measure the number of photons with energy $\hbar \omega_d$, then the eigenstates of the number operator of that mode, $N_d = b_d^\dag b_d$, would become the appropriate representation. The choice of the representation is not that clear nevertheless in the case of the quantum multiverse. After all, identifying the relevant physical observables in the multiverse is not at all an easy task. In the case of the quantum multiverse, it can be assumed that the Lewis number states, which are defined as the eigenstates of the invariant operator $I$, represent the occupation number for the quantum state of the multiverse, i.e. the number of universes in the multiverse, which is scale factor invariant.

However, other representations can be more useful in order to obtain the solutions of the Schr\"{o}dinger equation for the elements of the density matrix, i.e. $\frac{\partial }{\partial a} \langle \rho \rangle = \frac{i}{\hbar} \langle [H, \rho] \rangle$. For instance, let us consider the diagonal representation of the Hamiltonian (\ref{Hamiltonian 3-quantized}). It can be defined by the scale factor dependent creation and annihilation operators which correspond to the modes of the Hamiltonian with proper frequency, $\omega(a)$, i.e.
\begin{eqnarray}\label{annihilationomega}
b_\omega(a) & = & \sqrt{\frac{\omega(a)}{2 \hbar}} \left( \phi + \frac{i}{\omega(a)}  p_\phi \right) , \\  \label{creationomega}
b_\omega^\dag(a) & = & \sqrt{\frac{\omega(a)}{2 \hbar}} \left( \phi - \frac{i}{\omega(a)}  p_\phi \right) ,
\end{eqnarray}
where, $\omega(a) =\frac{\omega_0}{\hbar} a^{q-1}$. In such a representation, of course, the Hamiltonian turns out to be diagonal and given by, $H = \hbar \omega(a) (b^\dag_\omega b_\omega + \frac{1}{2})$. The annihilation and creation operators given by Eqs. (\ref{annihilationomega}) and (\ref{creationomega}) can, moreover, be related to the annihilation and creation operators of modes of unity frequency, $b_0$ and $b_0^\dag$, through the following squeezing transformation,
\begin{eqnarray}
b_\omega &=& \mu_\omega b_0 + \nu_\omega b_0^\dag , \\ b_\omega^\dag &=& \mu_\omega^* b_0^\dag + \nu_\omega^* b_0 ,
\end{eqnarray}
where,
\begin{eqnarray} \label{muomega}
\mu_\omega &=& \frac{1}{2} \left( \sqrt{\omega}+\frac{1}{\sqrt{\omega}} \right) , \\ \label{nuomega} \nu_\omega &=& \frac{1}{2} \left( \sqrt{\omega}-\frac{1}{\sqrt{\omega}} \right),
\end{eqnarray}
and hence, $\mu_\omega^2 - \nu_\omega^2 = 1$. Then, the new number states, $|N_\omega, a \rangle$, can be related to the number states of modes of unity frequency, $|N_0\rangle$, by
\begin{equation}\label{squeezednumber}
|N_\omega, a\rangle =  S_ \omega(\varepsilon_\omega)  |N_0 \rangle ,
\end{equation}
where the squeezing operator, $S_ \omega(\varepsilon_\omega)$, is defined by
\begin{equation}\label{Somega}
S_ \omega(\varepsilon_\omega) \equiv e^{\frac{1}{2}\varepsilon_\omega ( b_0^2 -  {b_0^\dag}^2 )}.
\end{equation}
$\varepsilon_\omega$ is the squeezing parameter which is related with Eqs. (\ref{muomega}) and (\ref{nuomega}) by, $\mu_\omega = \cosh \varepsilon_\omega$, and, $\nu_\omega = \sinh \varepsilon_\omega$. As a matter of completeness, let us note that the number eigenstates of the Hamiltonian can also be related to the Lewis number states, $| N, a\rangle$, defined in Sec. I by Eqs. (\ref{numberstates1})-(\ref{numberstates3}), as they are both linked to the number states of modes of unity frequency, $|N_0\rangle$. Thus, if $S_R(\varepsilon_R)$ is the squeezing operator that relates the Lewis number states, $| N, a\rangle$, with those of unity frequency, $|N_0\rangle$, defined by
\begin{equation}\label{SR}
S_R(\varepsilon_R) \equiv e^{\frac{1}{2}\varepsilon_R b_0^2 -  \frac{1}{2}\varepsilon_R^* {b_0^\dag}^2 } ,
\end{equation}
where, $\varepsilon_R = r e^{2 i \phi}$, with, $r = \cosh \mu_0$, and, $e^{2 i \phi} = \frac{\nu_0}{|\nu_0|}$, being $\mu_0$ and $\nu_0$ given by Eqs. (\ref{mu0}) and (\ref{nu0}), then $| N, a \rangle = S_R | N_0 \rangle$. Therefore, the eigenstates of the Hamiltonian, $|N_\omega\rangle$, are related to the number states, $|N,a\rangle$, through a more complicated squeezing transformation given by,
\begin{equation}\label{squeezednumber2}
|N_\omega, a\rangle =  S_ \omega(\varepsilon_\omega)  S_R^\dag(\varepsilon_R) |N, a \rangle .
\end{equation}
The squeezing transformations are unitary so that $S^\dag(\varepsilon) = S^{-1}(\varepsilon) = S(-\varepsilon)$, and therefore, they preserve the orthonormality relations among the number states of a given mode. Thus, the Hamiltonian eigenstates, $|N_\omega\rangle$, form also an orthonormal basis for the space of solutions. In order for these to satisfy the Schr\"{o}dinger equation, they have to be shifted by some phases which can be obtained following the same iterative, recurrent procedure as the one used in Ref. \cite{Lewis69}.

Then, the orthogonal representation of Hamiltonian eigenstates that satisfy the Schr\"{o}dinger equation is given by,
\begin{equation}
|\tilde{N}_\omega,a\rangle = e^{ i (N+\frac{1}{2}) \Omega(a) } |N_\omega,a\rangle ,
\end{equation}
where,
\begin{equation}\label{Omega}
\Omega(a) =  \int^a \omega(a') da' =  \frac{\omega_0}{\hbar q} a^q =  \frac{\omega(a)}{q} a .
\end{equation}
Let us now compute the scale factor dependence of the density matrix elements by solving the Schr\"{o}dinger equation in such a representation of Hamiltonian eigenstates, $|\tilde{N}_\omega,a\rangle $, in which the density matrix operator can be written as, $\rho = \sum_{N,M} P^\omega_{NM} |\tilde{N}_\omega, a\rangle \langle \tilde{M}_\omega, a |$. The solutions to the Schr\"{o}dinger equation are given then by,
\begin{equation}\label{Pnumber2}
P^\omega_{NM}(a) = e^{-i(M-N) \int^a da' \omega(a')} P^\omega_{NM}(0) ,
\end{equation}
with the diagonal elements taking on constant values. By employing another number representation it is possible to re-express the density matrix  in terms of squeezing transformation operators. For instance, in terms of the number states of modes of unity frequency, $|N_0\rangle$, the elements of the density matrix operator are given by,
\begin{eqnarray}
P_{I J}^0(a) & \equiv & \langle I_0 | \rho | J_0 \rangle \\ \nonumber & = & \sum_{N, M} P_{N M}^\omega(a) \langle I_0 | S_\omega(\varepsilon_\omega) | N_0 \rangle \langle M_0 | S_\omega^\dag (\varepsilon_\omega) | J_0 \rangle ,
\end{eqnarray}
where the involved matrix elements can no longer be trivially calculated.

\subsection{Coherent state representation}

Let us consider next the $P$ representation of the quantum state of the multiverse in terms of coherent states. These can be defined in terms of the number eigenstates of the Hamiltonian as,
\begin{equation}\label{cs}
| \alpha, a \rangle = e^{-\frac{|\alpha|^2}{2}} \sum_{N=0}^\infty \frac{\alpha^N}{\sqrt{N!}} e^{i (N+\frac{1}{2}) \Omega(a)} | N_\omega, a \rangle ,
\end{equation}
where, $\alpha = x + i y$, is a complex valued variable, and $\Omega(a)$ is given by Eq. (\ref{Omega}). The coherent states (\ref{cs}) are the eigenstates of the annihilation operator for a mode of frequency $\omega$, with scale factor eigenvalues given by,
\begin{equation}\label{annihilationP}
b_\omega(a) | \alpha, a \rangle = e^{i \Omega(a)} \alpha | \alpha, a \rangle .
\end{equation}
In the customary $P(\alpha)$ representation, the action of the creation operator upon the coherent state (\ref{cs}) is given by,
\begin{equation}\label{creationP}
b_\omega^\dag | \alpha, a \rangle = e^{- i \Omega(a)} \frac{\partial}{\partial \alpha} | \alpha, a \rangle ,
\end{equation}
and the density matrix can be written then as,
\begin{equation}\label{rhoP}
\rho(a) = \int d^2\alpha ||\alpha, a \rangle \langle \alpha, a || e^{- | \alpha |^2} P_\omega(\alpha, a) ,
\end{equation}
where, $||\alpha, a \rangle \equiv e^{ \frac{| \alpha |^2}{2}} |\alpha, a \rangle$, are the so called Bargmann coherent states \cite{Walls08}. Thus, using Eq. (\ref{rhoP}) in the Schr\"{o}dinger equation, $\partial_a \rho = \frac{i}{\hbar} [H, \rho]$, a differential equation is obtained for the function $P(\alpha, a)$, i.e.
\begin{equation}\label{schrodingerequation}
\frac{\partial}{\partial a} P_\omega(\alpha, a) = - i \omega(a) \left(  \alpha \frac{\partial}{\partial \alpha} - \alpha^* \frac{\partial}{\partial \alpha^*} \right) P_\omega(\alpha, a) .
\end{equation}
This equation can be easily solved by making the change, $\alpha = x + i y$. The solutions can be written as,
\begin{equation}\label{P}
P_\omega(\alpha, a) = C_1 P_1(\alpha, a) + C_2 P_2(\alpha, a).
\end{equation}
$C_1$ and $C_2$ are two constants to be determined by the boundary conditions and $P_1(\alpha, a)$ and $P_2(\alpha, a)$ are two independent solutions of the master equation (\ref{schrodingerequation}), given by
\begin{eqnarray}\label{P1}
P_1(\alpha, a) &=&  \frac{i}{2} \left( e^{-i \Omega(a)} \alpha - e^{i \Omega(a)} \alpha^* \right) , \\ \label{P2}
P_2(\alpha, a) &=&  \frac{1}{2} \left( e^{-i \Omega(a)} \alpha + e^{i \Omega(a)} \alpha^* \right) ,
\end{eqnarray}
where $\Omega(a)$ is once again given by Eq. (\ref{Omega}), and the integration constants have been chosen to ensure the real valued character of the function $P_\omega$. Eqs. (\ref{P1}) and (\ref{P2}) closely recall the form of the operator which represents the potential vector, $\vec{A}_{EM}(x,t)$, in the usual second quantization formalism of the electromagnetic field \cite{Glauber63}. Thus, Eq. (\ref{P}) may be interpreted as an oscillating field of a particular scale factor dependent mode, $\Omega(a)$, although the quantized field is now the wave function of a single universe.

Other solutions to Eq. (\ref{schrodingerequation}) can be found as well. For instance, by expanding in series the function $P_\omega$, i.e. $P_\omega = \sum_n (A_n \alpha^n + B_n {\alpha^*}^n) $, the solution can be written as,
\begin{equation}\nonumber
P_\omega(\alpha, a) =  \rm{Re}\left(  \frac{1}{1\pm e^{-i \Omega(a)} \alpha}  \right) P_\omega(0, 0) ,
\end{equation}
where $\Omega(a)$ is given by Eq. (\ref{Omega}), and the integration constants have been again chosen to ensure a real value for the function $P_\omega$.

In terms of other number representations, it may become difficult to carry out the explicit computation of the function $P$. For instance, let us consider a coherent state representation in terms of a general number representation,, i.e.
\begin{equation}\label{rhoP2}
\rho(a) = \int d^2\alpha  \, ||\alpha_g, a \rangle \langle \alpha_g, a || e^{- | \alpha |^2} P_g(\alpha, a) ,
\end{equation}
where,
\begin{equation}\label{csG}
|| \alpha_g, a \rangle =  \sum_{N=0}^\infty \frac{\alpha^N}{\sqrt{N!}} e^{i (N+\frac{1}{2}) \Omega_g(a)} | N_g, a \rangle.
\end{equation}
$| N_g, a \rangle$ are the number states of modes of a generic frequency, $\omega_g \equiv \omega_g(a)$, defined in terms of the annihilation and creation operators,
\begin{eqnarray}
b_g &=& \sqrt{\frac{\omega_g}{2 \hbar}} \left( \phi + \frac{i}{\omega_g} p_\phi \right) , \\ b_g^\dag &=& \sqrt{\frac{\omega_g^*}{2 \hbar}} \left( \phi - \frac{i}{\omega_g^*} p_\phi \right) ,
\end{eqnarray}
where for convenience we have retained the (plausible) complex character of the frequency of the mode. The phases in Eq. (\ref{csG}), which are proportional to $\Omega_g(a)$, account for the evolution of the number states like Schr\"{o}dinger vectors. Therefore, in such an unspecified representation the Schr\"{o}dinger states evolve into squeezed states, and the Hamiltonian is transformed into,
\begin{equation}
H = \hbar \left(  \gamma_- b_g^2 + \gamma_+ {b_g^\dag}^2  + \gamma_0 \left( 2 b_g^\dag b_g + 1 \right) \right) ,
\end{equation}
where $\gamma_\pm$ and $\gamma_0$ are functions which in general depend on the scale factor, except in the case of the diagonal representation of the Hamiltonian. They are given by,
\begin{eqnarray}
\gamma_-^* = \gamma_+ &=& \frac{|\omega_g|^2 \omega_g}{4 (\rm{Re}\omega_g)^2} \left( \frac{\omega^2}{\omega_g^2} - 1 \right) , \\ \gamma_0 &=& \frac{|\omega_g|^3}{4 (\rm{Re}\omega_g)^2} \left( \frac{\omega^2}{|\omega_g|^2} + 1 \right) ,
\end{eqnarray}
and therefore, $\Omega_g(a) = \int^a \gamma_0 da$. It can be checked that when the frequency of the representation, $\omega_g$, equals the proper frequency of the Hamiltonian, $\omega$, then such a representation diagonalizes the Hamiltonian. In general, the Schr\"{o}dinger equation for the density matrix, $\partial_a \rho = \frac{i}{\hbar} [H, \rho]$, can be written as,
\begin{eqnarray}
\int d^2\alpha ||\alpha_g\rangle \langle \alpha_g || e^{-|\alpha|^2} \partial_a P_g(\alpha, a) =  \;\;\;\;\;\;\;\;\;\;\;\;\;\;\;\;\;\;\;\;\;\;\;\;   \\ \;\;\;\;\;\;\;\; =\frac{i}{\hbar} \int d^2\alpha [H,||\alpha_g\rangle \langle \alpha_g ||] e^{-|\alpha|^2} \partial_a P_g(\alpha, a).
\end{eqnarray}
Taking into account that the relations (\ref{annihilationP}) and (\ref{creationP}) are satisfied now with $b_g$, $b_g^\dag$ and $\Omega_g(a)$ instead of $b_\omega$, $b_\omega^\dag$ and $\Omega(a)$, the differential equation for the function  $P_g(\alpha, a)$ can be written as,
\begin{equation}\label{general}
\frac{\partial}{\partial a} P_g(\alpha, a) = \left[ - \frac{\partial}{\partial x_j} A_j + \frac{1}{2} \frac{\partial}{\partial x_i}\frac{\partial}{\partial x_j} D_{i j} \right] P_g(\alpha, a) ,
\end{equation}
where, $(x_1, x_2) \equiv (\alpha, \alpha^*)$, the drift vector, $A_j$, is given by
\begin{equation}
A_j = (A_\alpha , A_{\alpha^*}) =( \xi_+ \, \alpha^* + i \gamma_0 \, \alpha ,  \xi_- \, \alpha - i \gamma_0 \, \alpha^*).
\end{equation}
The diffusion matrix, $D_{i j}$, is given by
\begin{equation}
D_{i j } = \left( \begin{array}{cc}
D_{\alpha \alpha} & 0  \\ 0 & D_{\alpha^* \alpha^*} \end{array} \right) = \left( \begin{array}{cc}
\xi_+ & 0  \\ 0 & \xi_- \end{array} \right) ,
\end{equation}
with, $\xi_-^* = \xi_+ = 2 i \gamma_+  e^{- 2 i \Omega_g}$, being, $\gamma_-^* = \gamma_+$. In general, Eq. (\ref{general}) is not easy to be analytically solved. However, before going any further it should be first clarified the meaning of the different representations for a particular cosmological model in terms of what can be observed, which is not at this stage clear at all.

\section{Comparative analysis with respect to other previous works}

The third quantization formalism has already been used in previous works on the multiverse and it has been criticised (see Ref. \cite{Vilenkin94}, and references therein). First of all, Vilenkin argued that an infinite number of fields and interaction types had to be introduced in order to describe topology changes in a four-dimensional realm, i.e. -- for all the non-simply connected space-times that can generally be considered.

Let us notice, however, that such a non-simply connected manifold can actually be converted into a simply-connected one by cutting it with a finite number of hypersurfaces \cite{Hawking90}, and this argument can be applied in the four-dimensional case. Hawking pointed out \cite{Hawking90} that the quantum state of the universe would then be given by a set of quantum oscillators, an idea similar to the multiverse scenario. In the third quantization formalism, on the other hand, the wave function which represents each of the disconnected regions might be given by that of a harmonic oscillator. In that case, the third quantization formalism adds a manageable procedure to study a multiverse made of disconnected regions, at least in the case of high symmetry.

The second main objection of Ref. \cite{Vilenkin94} was on the interpretation of the creation of universes in the third quantization formalism. Vilenkin argued that the usual interpretation in terms of $|in\rangle$ and $|out\rangle$ states, corresponding to $| N=0\rangle$ and $| N=1 \rangle$, respectively, with $N$ the number of universes, does not match the meaning of the creation of a universe from nothing proposed by himsef some years before \cite{Vilenkin84}. The  $| in\rangle$ state would correspond to the "nothing state" when the scale factor decreases, whereas the $|out \rangle$ state would describe a single universe with a large value of the scale factor, i.e. when $a \rightarrow \infty$. However, as Vilenkin himself objected \cite{Vilenkin94}, the scale factor cannot be used as a time variable because it is not a monotonic variable. Moreover, taking a particular value of the scale factor is nothing but considering a particular time reparametrization, $a=a(t)$, and it is not expected that the number of universes in the whole multiverse would depend on a particular choice of a reference system within a single universe.

On the other hand, we have shown that the number state which properly defines the number of universes is represented by the Lewis number state defined by Eq. (\ref{numberstates3}) and thus, the number of universes in the whole multiverse is scale factor invariant, i.e. $N \neq N(a)$. However, the concept of universes created from nothing can still be meaningfully kept because the vacuum state of the multiverse, i.e. the \emph{void}, still represents the absence of space, time and matter, and is what Vilenkin means when he uses the term "nothing".

Finally, it was considered in Ref. \cite{Hawking06} that in the context of eternal inflation a mosaic structure made up of causally disconnected thermalized regions should appear with different values of the effective coupling constants, with a probability distribution for their values. However, it is not at all clear that the general state of the multiverse has to be a thermal state because, unlike the inflationary model, a common space-time is not guaranteed at all in the case of a general multiverse.

In our simplified model, on the other hand, a vanishing cosmological constant seems probable, in agreement with the result of Ref. \cite{Coleman88b}. However, there may be a solution to this problem, because we have shown that in terms of the creation and annihilation operators for single universes, $b^\dag$ and $b$, a non-zero value of the vacuum energy should be ascribed to each universe, even at $a = 0$. Therefore, although the most probable value for the cosmological constant would be zero in the realm of the whole multiverse, if it is taken to be formed up by pairs of universes, a non-zero value for the vacuum energy could be present in each single universe due to entanglement with its partner, which could be responsible both for the currently observed accelerating expansion and for the value of the vacuum energy in the inflationary era.

\section{Conclusions and further comments}

In this paper, we have applied a third quantization formalism to
the case of a multiverse made up of homogeneous and isotropic space-times filled with
a perfect fluid. A well-defined quantum state
of such a multiverse has been obtained in terms of the states of a
harmonic oscillator, and it satisfies the most familiar boundary
conditions which are used in quantum cosmology. It can also be given in
terms of a wavefunction or a density matrix, accounting for pure and mixed
states. Thus, the third quantization scheme can be completely well-defined in these models of high symmetry.

The state of the multiverse turns out to be given by a squeezed
state with no classical analog. On the one hand, the quantum state
for a gravitational vacuum which is filled with a large number of
popping baby universes and wormholes connecting them, is given by
a squeezed state in which quantum correlations among the number
states are dominant. This result agrees with that of Ref. \cite{Grishchuk90}, in
which the vacuum state of gravity is populated with relic
gravitons which unavoidably evolve also into a highly squeezed
state having no classical analog. On the other hand, in the case of large
parent universes the quantum correlations among the number states
 disappear asymptotically, the number state representation turns
out to be an appropriate one and the single universe approximation
can be then considered. Thus, once a single parent universe can be
assumed, quantum transitions to larger number states are
asymptotically suppressed. Notwithstanding, if universes are created in pairs, then the correlations in the quantum state of a pair of universes would imply that each single universe is not independent. Then, one of the main non-independence consequences might be the presence of an entanglement energy, which could account for both the large value of the cosmological constant in the inflationary period and for the small value of the vacuum energy now, because the entanglement between the pair of universes decreases as each single universe expands.

Finally, let us notice that the model presented in this paper can
be naturally extended to include local matter fields as well as
cosmological fields. In the first case, it is expected that their
effect would represent a minor correction to the value of the
frequency of the harmonic oscillator which quantum mechanically
described the multiverse. However, in the latter case, derivatives
with respect to the matter fields appear in the Wheeler-De Witt
equation, implying that the state of the multiverse is given
in terms of the modes of a wave equation in which the mass and the frequency might be scale factor dependent. Nevertheless, it seems that it will be easier to work with the solutions of such a wave equation rather
than with the general solutions of the second quantized theory.

\acknowledgements The authors want to thank A. Forrester for his
invaluable help in the formal elaboration of this paper. This work
was supported by MICINN under research project no. FIS2008-06332.

\end{document}